\documentclass[journal]{IEEEtran}

\usepackage{graphicx,epsfig}
\usepackage{amsmath,mathrsfs} 
\usepackage{algorithm}
\usepackage[noend]{algpseudocode}
\makeatletter
\def\BState{\State\hskip-\ALG@thistlm}
\makeatother
\usepackage{amssymb} 
\usepackage{amsthm}
\usepackage{amsfonts}
\usepackage{mathtools}
\usepackage{multirow}
\usepackage{mathtools} 
\usepackage{cite}
\usepackage{epstopdf}
\usepackage{ifthen}
\usepackage{bbm}
\usepackage{setspace}

\newcommand{\C}{ \mathbb{C}}

\def\Spc#1{{\mathcal{#1}}}  
\def\M#1{{\bf{#1}}}  

\makeatletter
\DeclareRobustCommand\onedot{\futurelet\@let@token\@onedot}
\def\@onedot{\ifx\@let@token.\else.\null\fi\xspace}

\makeatother

\usepackage{siunitx}

\def \upomega{\mathbf{\omega}}
\def \uppi{\mathbf{\pi}}

\usepackage{setspace}   
\usepackage{lipsum} 

\usepackage{graphicx,epsfig}
\usepackage{amsmath,mathrsfs} 
\usepackage{algorithm}
\usepackage[noend]{algpseudocode}
\makeatletter
\def\BState{\State\hskip-\ALG@thistlm}
\makeatother
\usepackage{amssymb} 
\usepackage{amsthm}
\usepackage{amsfonts}
\usepackage{mathtools}
\usepackage{multirow}
\usepackage{mathtools} 
\usepackage{cite}
\usepackage{epstopdf}
\usepackage{ifthen}
\usepackage{bbm}
\usepackage{setspace}
\usepackage{color}
\newcommand{\arch}[1]{\textsc{#1}}
\newcommand{\Fref}[1]{Figure \ref{#1}}

\newcommand{\Tref}[1]{Table \ref{#1}}
\renewcommand{\S}{Section }

\usepackage{lipsum}
\usepackage{blindtext}
\usepackage{kotex}
\usepackage{xcolor}

\begin{document}
\title{Time-Dependent Deep Image Prior for Dynamic MRI}

\author{Jaejun Yoo,~Kyong~Hwan~Jin,~Harshit~Gupta,~J{\'{e}}r{\^{o}}me~Yerly,~Matthias~Stuber,
        and~Michael~Unser,~\IEEEmembership{Fellow,~IEEE}
        
 \thanks{
 J. Yoo, K.H. Jin, H. Gupta, and M. Unser  are with Biomedical Imaging Group, \'{E}cole polytechnique f{\'{e}}d{\'{e}}rale de Lausanne (EPFL), Switzerland. (e-mail: \mbox{jaejun.yoo88@gmail.com}; {kyonghwan.jin@gmail.com}; {harshit.gupta.cor@gmail.com}; michael.unser@epfl.ch)
 
 J. Yerly  and M. Stuber are with Department of Radiology, University Hospital (CHUV) and University of Lausanne (UNIL), and Center for Biomedical Imaging (CIBM), Lausanne, Switzerland. (e-mail: \mbox{yerlyj.mri@gmail.com};~matthias.stuber@chuv.ch )
 
 This work has been funded partly by European Research Council (ERC) Grant 692726 (H2020-ERC Project GlobalBioIm). (Corresponding author: J. Yoo) 
 }
}

\maketitle

\begin{abstract}
We propose a novel unsupervised deep-learning-based algorithm for dynamic magnetic resonance imaging (MRI) reconstruction. Dynamic MRI requires rapid data acquisition for the study of moving organs such as the heart. Existing reconstruction methods suffer from restrictions either in the model design or in the absence of ground-truth data, resulting in low image quality. We introduce a generalized version of the deep-image-prior approach, which optimizes the network weights to fit a sequence of sparsely acquired dynamic MRI measurements. Our method needs neither prior training nor additional data. 
In particular, for cardiac images, it does not require the marking of heartbeats or the reordering of spokes.  The key ingredients of our method are threefold: 1) a fixed low-dimensional manifold that encodes the temporal variations of images; 2) a network that maps the manifold into a more expressive latent space; and 3) a convolutional neural network that generates a dynamic series of MRI images from the latent variables and that favors their consistency with the measurements in k-space. 
Our method outperforms the state-of-the-art methods quantitatively and qualitatively in both retrospective and real fetal cardiac datasets. 
To the best of our knowledge, this is the first unsupervised deep-learning-based method that can reconstruct the continuous variation of dynamic MRI sequences with high spatial resolution. 

\end{abstract}

\section{Introduction}\label{sec: Introduction}
The aim of dynamic magnetic resonance imaging (MRI) is to capture the dynamics associated with moving organs, which requires a fast imaging process. A typical approach is to accelerate data acquisition by a partial sampling of the k-space. The resulting partial loss of data must then be compensated to maintain the image quality. 
Several methods have addressed this by exploiting spatial or temporal redundancy, including parallel MRI~\cite{griswold2002generalized,pruessmann1999sense, kellman2001adaptive, breuer2005dynamic}, k-t acceleration methods \cite{tsao2003k, huang2005k,xu2007improving}, compressed sensing (CS) MRI~\cite{jung2007improved, gamper2008compressed,ji2008dynamic, jung2009k, otazo2010combination, wang2013compressed,feng2013highly, feng2014golden, feng2016xd}, low‐rank methods \cite{lingala2011accelerated,poddar2015dynamic,otazo2015low,nakarmi2017ker,nakarmi2017m,nakarmi2018mls, ravishankar2017low}, and many others. 
In the specific case of cardiac applications, the current state-of-the-art methods further improve the reconstruction by exploiting the fact that the heart motion is approximately cyclic. They typically use electrocardiograms or self-gating techniques~\cite{yerly2016coronary,chaptinel2017fetal}. However, all of these methods are limited by constraints over the signal-to-noise ratios (SNR), restrictions in the coil design, hand-picked priors, multiple processing steps, or inefficient algorithms in their deployment of the standard convex-optimization techniques. 

More recently, inspired by the development of deep-learning techniques in various imaging modalities~\cite{gupta2018cnn,kang2018deep,yoo2018mathematical, yoo2019deep}, supervised-learning approaches have been applied to the fast and accurate reconstruction of partially sampled MRI~\cite{sun2016deep,wang2016accelerating,jin2017deep,tezcan2018mr, hammernik2018learning,han2018deep,schlemper2018deep,hauptmann2019real,mardani2019deep}. These methods, however, heavily depend on a training dataset, especially on  ground-truth data (\textit{i.e.}, fully sampled measurements), which are typically unavailable for dynamic MRI. Unlike the direct deep-learning approaches, the model-based deep-learning framework of  \cite{biswas2019dynamic} formulates the image recovery as an optimization scheme. By unrolling an iterative algorithm, it minimizes a cost function that combines data consistency and a deep-learned prior. Because the learned prior incorporates patient-specific noise patterns into the algorithm, this approach successfully recovers images with fast reconstruction and acceptable quality. However, it still requires ground-truth data to train the denoising network. 
\subsection{Contribution}\label{sec: Contribution}
In this paper, we propose an unsupervised learning framework in which a generative network is optimized to reconstruct a sequence of golden-angle radial lines in k-space, also called spokes. 
Inspired by deep image priors (DIP)~\cite{ulyanov2018deep}, we use a convolutional neural networks (CNN) architecture as an implicit structural prior that constrains the search space of the optimization problem. In addition, to learn the temporal dependencies of the dynamic measurements, we impose a one-dimensional manifold parameterized by time. Aided by this explicit cue, the network then learns to encode the temporal variations of the sequential images into the spatial closeness of the samples on the imposed manifold. This simple temporal coupling already enables our model to outperform the other CS algorithms~\cite{feng2014golden,feng2016xd, yerly2016coronary,chaptinel2017fetal} without bells and whistles---note that our approach is purely unsupervised and optimized in an end-to-end manner. We further improve the reconstruction by introducing a mapping network (MapNet) that brings more flexibility to our latent space \cite{karras2019style, choi2020stargan}. MapNet consists of a few fully connected layers with nonlinear activations; it learns to map the fixed manifold into a more expressive latent space. This allows the subsequent generative network to adapt its input to a given dataset, thereby improving image quality (\Fref{fig:overview}). 


In short, our generative model takes the latent variables from MapNet and reconstructs dynamic images by exploiting its powerful structural prior. 
With the extensive analyses in \S\ref{sec: Analysis} and experimental results in \S\ref{sec: Results}, we show that both the manifold design and MapNet are essential to achieve good reconstructions. To the best of our knowledge, this is the first unsupervised deep-learning-based method that can reconstruct the full temporal frames of dynamic MRI sequences with high spatial resolution.
\subsection{Related Work}\label{sec: Related Work}
\medskip
\noindent\textbf{Unsupervised Learning.} 
Starting from the seminal work of DIP~\cite{ulyanov2018deep}, there have been several studies that applied unsupervised learning to medical imaging, such as MRI \cite{yazdanpanah2019non} and positron emission tomography \cite{gong2018pet}, albeit both cases address the reconstruction of static images. 

The closest work to ours is the one that used DIP for video compression \cite{hyder2019generative}, which also considers a sequence of latent inputs. However, unlike our goal (the reconstruction of an image sequence), theirs is to find compact codes for the representation of video frames. 
To find such codes, they optimize both the network weights and latent variables. Without any constraint on the latent space, however, the latent codes may diverge to an arbitrary space. To prevent this, they imposed either low-rank or similarity constraints on the latent sequence. 
However, this optimization not only requires additional effort to tune hyper-parameters but also entails a singular-value decomposition at each iteration, which severely increases the computational burden. 
By contrast, we solve this by simply inputting an explicit manifold and letting a mapping network adapt the manifold to the given data. This makes the training much easier and yields the one-dimensional manifold in latent space that is adapted to the given data. 
In addition, their forward model is an identity operator, while ours is an MR measurement operator with severe under-sampling. 

\section{Methods}\label{sec: Methods}

\begin{figure*}
\begin{center}
\includegraphics[trim = 0mm 0mm 0mm 0mm,clip=true, width=\linewidth]{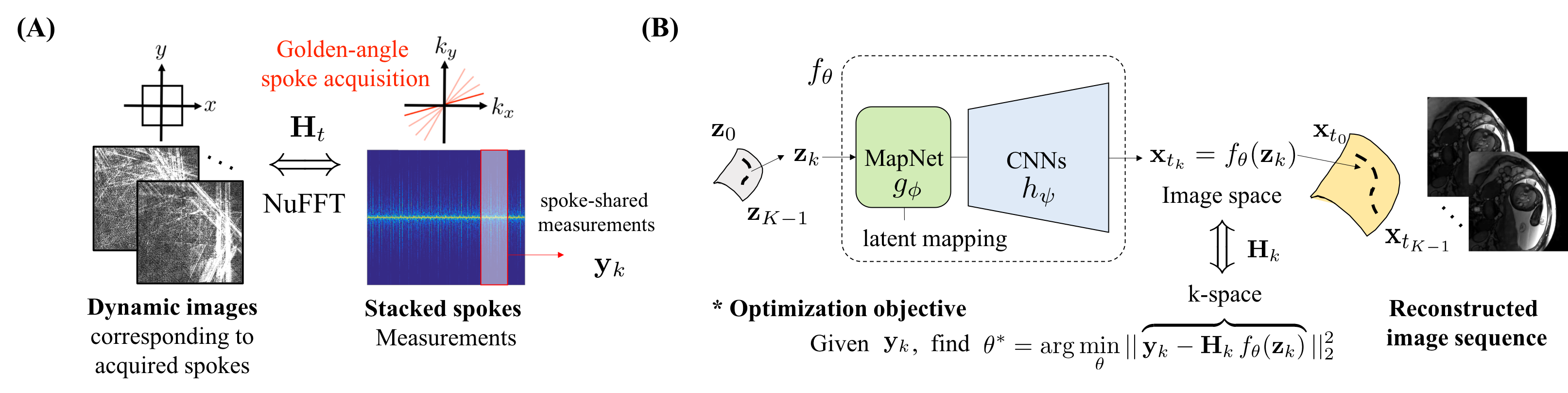}
\caption{\label{fig:overview} Overview of our framework. (A) Schematic illustration of the dynamic MRI data-acquisition procedure. We use a nonuniform fast Fourier transform with a golden-angle scheme and spoke sharing. (B) Proposed framework based on latent mapping and deep image priors. The block labeled $g_\phi$ is a mapping network (MapNet) of fully connected layers and the block labeled $h_\psi$ denotes the generative CNNs. 
}
\end{center}
\end{figure*}




We first briefly recapitulate the content of deep image prior (\S\ref{sec:dip}) as well as the physics of dynamic MRI (\S\ref{sec:dynamicmri}). Then, we describe our method based on DIP with a mapping network and on the learning of the underlying latent manifolds 
(\S\ref{sec:proposed}). 

\subsection{Deep Image Prior}\label{sec:dip}
The deep image prior \cite{ulyanov2018deep} is a recent approach that has been proposed for solving static linear inverse problems, such as image denoising, inpainting, and superresolution. 
DIP has been found to capture advanced image statistics in a purely unsupervised way by using a strong structural prior of convolutional neural networks, with neither any prior training nor additional image data. 
Taking a random but fixed latent variable $\M z \in \mathbb{R}^L$ as an input, DIP optimizes the parameters $\boldsymbol \theta$ of an untrained neural network $f_{\boldsymbol\theta}$ to produce an output $f_{\boldsymbol\theta}(\M z)$ that is consistent with the measurement $\M y \in \mathbb{R}^M$. The problem being solved is formalized as  
\begin{equation}
\label{eq:dip}\boldsymbol\theta^{*}=\arg\min_{\boldsymbol\theta}\left\|{\mathbf{y}}-\M H(f_{\boldsymbol\theta}({\mathbf{z}}))\right\|_{2}^{2},
\end{equation}
where $\M H\in\mathbb{R}^{M\times N}$ is a forward model. For example, in the image superresolution problem, $\M y$ is a noisy low-resolution image and $\M H$ is a downsampling operator. 
The output of the optimized network $\M x^*=f_{\boldsymbol\theta^*}({\mathbf{z}})$ then yields a reconstructed image of surprisingly good quality. 
This has been ascribed to the implicit representation bias of the CNN architecture, which favors a natural-looking output image over a noisy unstructured one. In this paper, we extend the concept of DIP to solve a more challenging dynamic MRI reconstruction problem. 


\subsection{Dynamic MRI}\label{sec:dynamicmri}
We use a radial 2D MRI acquisition scheme where the instrumentation is such that it physically records a temporal sequence of radial lines of the Fourier transform of a fixed slice (image) of a 3D volumetric object. The underlying 2D image is represented by a vector ${\mathbf{x}}\in{\mathbb{C}}^{N}$, where $N$ is the number of pixels. 
At a given time point $t$, the vector of $k$-space measurements ${\mathbf{y}}(t)\in{\mathbb{C}}^{M_0}$ consists of the uniform samples of the 2D Fourier transform of the image taken along a radial line at some orientation $\vartheta=\vartheta(t)$. Because of the central-slice theorem, these measurements can also be interpreted as the 1D Fourier transform of the Radon transform of the image at angle $\vartheta$.
By repeating this process with a sufficiently dense sequence of angles $\vartheta_k \in[0,\pi)$, and assuming the images to be static, one  obtains a complete data set from which a high-quality (static) image can be reconstructed using standard tomographic techniques. 
Now, the difficulty with dynamic imaging is that the underlying image is not static 
but varies through time, which calls for a more sophisticated reconstruction procedure.
\subsubsection{Forward Model}
The measurement process that relates the image at time $t$ and the $k$-space measurements with angle $\vartheta=\vartheta(t)$ is linear and formally described by the relation
\begin{equation}
\label{eq:forward static}
\M y(t)=\M H\big(\vartheta\big) \M x(t),     
\end{equation}
where $\M H(\vartheta)$ is the $M_0 \times N$ system matrix that represents the combined effect of taking the 2D Fourier transform of $\M x$ and resampling along a radial line with direction $\vartheta$. The type of measurement provided by \eqref{eq:forward static} is referred to as an angular {\em spoke}.
In practice, we acquire a series of $K$ spokes taken at regularly spaced time point $t_k=t_0+k \Delta t$, $k=0,\dots, (K-1)$ with step size $\Delta t$. The spoke orientations follow the golden-angle strategy 
\begin{equation}
\label{eq: spoke-sharing}\vartheta_{k}=\vartheta_{0}+\omega_{0}\,k\Delta t,
\end{equation}
where $\vartheta_{k}$ gives the orientation of a spoke at time $t_k = t_0+k\Delta t$, with $\omega_{0}$ its angular velocity. The golden-angle specificity is the irrationality condition $\left(\omega_{0}\,\Delta t/\uppi\right)\notin{\mathbb{Q}}$, which is approximated by setting $\left(\omega_{0}\,\Delta t\right)\approx111.25^{\circ}$~\cite{feng2014golden}. 
Then, our task is to reconstruct the image sequence $\{\M x(t_k)\}_{k=0}^{K-1}$ from the measurement sequence $\{\M y(t_k)\}_{k=0}^{K-1}$. 

\subsubsection{Spoke-Sharing}\label{sec: spoke-sharing}
The ambitious goal of accelerated dynamic MRI is to reconstruct $\{\M x (t_{k})\}_{k=0}^{K-1}$---or, even better, ${\mathbf{x}}(t)$ for $t\in[t_0,T_{K-1}]$---from the finite set of measurements $\{{\mathbf{y}}(t_{k})\}$. However, a single orientation per frame does not provide enough information to recover the corresponding instantaneous two-dimensional image ${\mathbf{x}}(t_{k})$. 
To overcome this issue, we assume that the changes are slow over some small number of neighboring spokes ($n_{\rm s}$), so that ${\mathbf{x}}(t)\approx{\mathbf{x}}(t_{k})$ for all $t \in {\mathbb{T}}_{{k}}=[t_{k}-n_{\rm s}\,\Delta t/2,t_{k}+n_{\rm s}\,\Delta t/2)$. The sharing parameter $n_{\rm s}\in2\,{\mathbb{N}}+1$ is the number of radial lines used for the reconstruction of one frame; it controls the temporal resolution.

To further describe this pooling process, we introduce the augmented
measurement vector ${\mathbf{y}}_{{k}}=\big(\mathbf{y}(t_m)\big)_{m=k-(n_{\rm s}-1)/2}^{k+(n_{\rm s}-1)/2}$ of size $M=n_{\rm s} \times M_0$.
Correspondingly, we define the column-wise concatenated system matrix
${\mathbf{H}}_{{k}}=({\mathbf{H}}(\vartheta_k))_{m=k-(n_{\rm s}-1)/2}^{k+(n_{\rm s}-1)/2}$, whose time dependence is indicated by the index $k$.
This results in the forward imaging model
\begin{equation}
\label{eq:forward}{\mathbf{y}}_{{k}}={\mathbf{H}}_{{k}}\,{\mathbf{x}}(t_{k}),
\end{equation}
where the matrix ${\mathbf{H}}_{{k}} \in \C^{M \times N}$ encodes the (pseudo-simultaneous) acquisition of $n_{\rm s}$ spokes at time $t_k$.
The underlying strategy is called spoke-sharing. Because of the irrationality condition of the golden-angle approach, no direction will ever be measured twice.
While the imaging model \eqref{eq:forward} is more favorable than \eqref{eq:forward static} because of the augmented number of measurements,
the problem is still ill-posed because  $M=n_{\rm s}\,M_{0}$ remains smaller than $N$ (the number of unknowns).
The common practice, therefore, is to introduce an appropriate regularizer. In this paper, we propose to constrain the solution by applying a deep image prior that is shared among all frames.  

\subsection{Proposed Framework} \label{sec:proposed} 
To address the dynamic MRI reconstruction problem, we first modify the original DIP so that it takes a sequence of input and output pairs (\Fref{fig:overview}). 
More specifically, we optimize an untrained neural network $f_{\boldsymbol\theta}$ to map a sequence of inputs $\{\M z_k\}_{k=0}^{K-1}$ to the spoke-shared measurements $\{\M y_k\}_{k=0}^{K-1}$, thereby reconstructing the sequence of images $\{\M x(t_k)\}_{k=0}^{K-1}$ by searching for
\begin{align}
    \boldsymbol\theta^*&=\arg \min_{\boldsymbol\theta}\frac{1}{K}\sum_{k=0}^{K-1}\|\M y_k-\M H_k  f_{\boldsymbol\theta}(\M z_k)\|^2, \label{eq:ours} 
\end{align}
leading to $\M x^*(t_k)=f_{\boldsymbol\theta^*}(\M z_k)$. 
Note that the optimization is done in the measurement domain. 
This 
enforces the image sequence to be consistent with the measurements, while the modified DIP scheme regularizes the reconstructed images. 


\medskip
\noindent\textbf{Manifold Design.} 
To fully exploit the characteristics of dynamic MRI, the underlying model must be able to effectively encode the temporal variations of the measurements while preserving the structure of the individual frames. 
To this end, we propose to design a manifold $\mathcal{Z}$, thereby effectively injecting a specific prior into the network. 
For example, an ordered sequence $\{\M z_k\}$ from a straight-line manifold will guide the network to associate spatial closeness of input variables with temporal closeness of images. This encourages the network to reconstruct an image sequence with temporally similar attributes. For a quasi-periodic signal such as the cardiac motion, we can encode the expected behavior by letting the manifold take the structure of a 
three-dimensional helix. 

\medskip
\noindent\textbf{Mapping Network (MapNet).} 
Although a careful choice of temporally meaningful manifolds typically results in an excellent performance, the fact that the design is hand-crafted may also sometimes limit the performance of the network \cite{karras2019style}. 
To add flexibility to our model and to exploit the rich representation power of the network, we introduce a mapping network (MapNet). In our design, MapNet $g_\phi$ involves a few fully connected layers with nonlinearities. It learns to map a fixed manifold into the more expressive latent space $\mathcal{W} = g_\phi(\mathcal{Z})$. 
More specifically, 
our model $f_{\boldsymbol\theta}$ now has a hierarchical architecture that consists of the MapNet $g_\phi$ followed by CNN $h_\psi$ so that $f_{\boldsymbol\theta}=h_\psi \circ g_\phi$ and $\boldsymbol\theta=\{\phi, \psi\}$ (\Fref{fig:overview} (B)). This leads us to replace \eqref{eq:ours} by 
\begin{align}
L_K(\boldsymbol\theta)=\frac{1}{K}\sum_{k=0}^{K-1}\|\M y_{k} -\M H_{k} (h\circ g)_{\boldsymbol\theta}(\M z_{k})\|^2.
\label{eq:ours_final}
\end{align}
The role of $g_\phi$ is to appropriately warp the input manifold to facilitate $h_\psi$ in its reconstruction of the true dynamics. Overall, the insertion of $g_\phi$ provides better flexibility to our model and lets us efficiently exploit the representation power of neural networks, resulting in a good reconstruction. 

\medskip
\noindent\textbf{Final Algorithm.} Our optimization scheme is given in Algorithm \ref{DMUNN-algorithm}. 
We minimize the loss function \eqref{eq:ours_final} using standard gradient-descent methods \cite{kingma2014adam} for $n_{\rm iter}$ iterations. At each iteration, instead of \eqref{eq:ours_final}, a batch loss $L_{B}(\boldsymbol\theta)$ is updated where a batch $\{k_0,\ldots,k_{B-1} \}$ of size $B$ is randomly sampled from the index set $\{0, \ldots, K-1\}$. The corresponding input variables $\{\M z_{k_b}\}_{b=0}^{B-1}$ are fed to the network and 
its parameters are updated using the gradient with respect to $\boldsymbol\theta$. 
\begin{table*}[ht]
\begin{center}
\begin{tabular}{cccccc}\hline\hline
Operation Layer&\parbox{0.75 in}{\begin{center}Number of Filters\end{center}}&\parbox{0.75in}{\begin{center}Size of Each Filter (XYC)\end{center}}&\parbox{0.5in}{\begin{center}Strides (XY)\end{center}}&\parbox{1 in}{\begin{center}Zero Padding (XY)\end{center}}&\parbox{1 in}{\begin{center}Size of Output Image (XYC)\end{center}}\\
\hline 
Input of $h_\psi$ ($L=64$)&&&&&$8\times8\times1$ \\
Conv+BN+ReLU&$128$&$3\times3\times1$&$1\times1$&$1\times1$&$8\times8\times128$\\
Conv+BN+ReLU&$128$&$3\times3\times128$&$1\times1$&$1\times1$&$8\times8\times128$\\
NN interp.&&&$2\times2$&&$16\times16\times128$\\
2$\times$(Conv+BN+ReLU)&$128$&$3\times3\times128$&$1\times1$&$1\times1$&$16\times16\times128$\\
NN interp.&&&$2\times2$&&$32\times32\times128$\\
2$\times$(Conv+BN+ReLU)&$128$&$3\times3\times128$&$1\times1$&$1\times1$&$32\times32\times128$\\
NN interp.&&&$2\times2$&&$64\times64\times128$\\
2$\times$(Conv+BN+ReLU)&$128$&$3\times3\times128$&$1\times1$&$1\times1$&$64\times64\times128$\\
NN interp.&&&$2\times2$&&$128\times128\times128$\\
2$\times$(Conv+BN+ReLU)&$128$&$3\times3\times128$&$1\times1$&$1\times1$&$128\times128\times128$\\
Conv.&$2$&$3\times3\times128$&$1\times1$&$1\times1$&$128\times128\times2$\\
\hline\hline
\end{tabular}
\end{center}
\caption{\label{tab:arch}Architecture of our generative convolutional network ($h_\psi$). Conv.: convolution; BN: batch normalization; NN interp.: nearest-neighbor interpolation.}
\end{table*}

\begin{algorithm}[t]
	\caption{Time-dependent DIP for dynamic MRI. We use Adam optimizer \cite{kingma2014adam} with $n_{\rm iter}=10,000$ and  $B=1$.}
	\label{DMUNN-algorithm}
\textbf{Input}: Set of measurements $\{\M y_k\}_{k=0}^{K-1}$, number of iterations $n_{\rm iter}$, batch size $B$, and number of cycles $p$. 
\begin{enumerate}
    \item Select a manifold $\mathcal{Z}$.
    \item Sample $\{\M z_k\}_{k=0}^{K-1}$ from $\mathcal{Z}$.

\item Optimize $\boldsymbol\theta$.\\
\textbf{for} $n_{\rm iter}$ iterations \textbf{do}
\begin{itemize}[\leftmargin=1 in]
    \item Randomly sample a batch  $\{k_0,\ldots,k_{B-1} \}$ of size $B$ from $\{0, \ldots, K-1\}$.
    \item Compute the batch loss of \eqref{eq:ours_final}.
    \item Update $\boldsymbol\theta$ with gradient $\nabla_{\boldsymbol\theta}L_{B}(\boldsymbol\theta)$.
\end{itemize}
\textbf{end} \textbf{for}

 \item Reconstruct images $\left\{\left(h\circ g\right)_{\boldsymbol\theta^*}(\M z_k)\right\}_{k=0}^{K-1}$.
\end{enumerate}
\end{algorithm}

{\section{Experiments}}\label{sec: Architectures, Datasets, and Training}
In this section, we describe the datasets, baseline methods, cardiac-cycle estimation, evaluation setups, and implementation details. 

\subsection{Datasets}
All experimental datasets are breath-hold MR images. We assume a twofold upsampling of measurements for every dataset. Therefore, the size of the reconstructed fields of view is half that of the first dimension of the measurements.

\subsubsection{Retrospective Dataset} A cardiac cine dataset was acquired using a 3T whole-body MRI scanner (Siemens; Tim Trio) equipped with a 32-element cardiac coil array. The acquisition sequence was bSSFP and prospective cardiac gating was used. The imaging parameters were as follows: FOV=$(300\times300)\,\mbox{mm}^{2}$, acquisition matrix size=$(128\times128)$, TE/TR=$1.37/2.7\,\mbox{ms}$, receiver bandwidth=$1184\,\mbox{Hz/pixel}$, and flip angle=$40^{\circ}$. The number of frames was $23$ and the temporal resolution was $43.2\,\mbox{ms}$. The resulting fully sampled Cartesian trajectories are used as ground-truth. To retrospectively simulate the radial sampling, we implemented the forward model using the golden-angle strategy with NuFFT\footnote{https://github.com/marchdf/python-nufft}. Sinograms are obtained as shown in Figure~\ref{fig:overview}.  The number of spokes per frame is $n_s=13$. For a single-cycle simulation, 
the dimension of sinograms is $(K\times n_s \times  M_{\upomega}\times C)=(23\times13\times256\times32)$. For a multicycle simulation, we acquire $p=13$ cycles, which results in $K=13\cdot23=299$ frames.  

\subsubsection{Fetal Cardiac Dataset}
Fetal cardiac MRI data were acquired on a 1.5 T clinical MR scanner (MAGNETOM Aera, Siemens AG, Healthcare Sector, Erlangen, Germany) with an 18-channel body array coil and a 32-channel spine coil for signal reception. We used an untriggered continuous 2D bSSFP sequence that was modified to acquire radial readouts with a golden-angle trajectory \cite{chaptinel2017fetal}. The acquisition parameters were: FOV = (260 $\times$ 260) $\text{mm}^2$, acquisition matrix size = (256 $\times$ 256) pixels, slice thickness = 4.0 mm, TE/TR = 1.99/4.1 ms, RF excitation angle = 70$^\circ$, radial readouts = 1400, acquisition time = 6.7 s, and bandwidth = 1028 Hz/pixel.

\subsection{Baseline Methods}
We apply three baseline methods. 
\begin{enumerate}
 \item \textbf{Back Projection (BP)} is a zero-filled discrete Fourier transform, which is the most basic baseline one can think of.  
 \item   \textbf{GRASP~\cite{feng2014golden}} is a golden-angle radial sparse parallel MRI algorithm, which extends the idea of k-t SPARSE-SENSE \cite{otazo2010combination} to volumetric golden-angle radial acquisitions. Here, the spoke-sharing strategy is not applied.
 \item   \textbf{Reordering Method (RD)~\cite{yerly2016coronary,chaptinel2017fetal}} is a three-step algorithm. RD first reconstructs real-time images of limited image quality and uses these images to reorder or self-gate the measurements, which in turn are used for the final reconstruction with k-t SPARSE-SENSE \cite{otazo2010combination}. In the retrospective experiment, where we know the phase indices, we use the exact order of frames for self-gating. 
\end{enumerate}
 
\subsection{Estimation of cardiac Cycles}
For the processing of the fetal cardiac dataset, RD and our algorithm both require a rough estimate of the number of cardiac cycles seen over the whole duration of a sequence of data acquisition. It can be typically obtained from k-space. 
Simple techniques to estimate the cardiac cycles from radial data have been previously reported by \cite{di2019automated, yerly2016coronary,larson2004self}. Radial acquisition schemes sample the center of k-space at every readout, which supports the extraction of physiological motion signals. The central k-space coefficient of a radial readout (\textit{i.e.}, the echo peak) corresponds to the complex sum of the transverse magnetization across the entire image volume. In the presence of moving structures such as a beating heart, changes in the overall transverse magnetization due to motion will induce a modulation of the consecutive echo peaks. (Trajectory imperfections and eddy currents can also modulate echo peaks, but their frequency responses differ from the physiological motion frequencies and, thus, can be filtered out.) The resulting signal can then be used to estimate the number of cardiac cycles and to inform the manifold network. For our fetal cardiac dataset, we find that the time-course has approximately 13 periods so that we finally set $p=13$. 

\subsection{Evaluation Metric}
We use the regressed SNR as a quantitative metric. With the oracle $\mathbf{x}$ and the reconstructed image ${\mathbf{x}}^{*}$, RSNR is given by 
\begin{equation}
\mbox{RSNR}=\max_{a,b\in\mathbb{R}}20\,\log\,\frac{\left\|{\mathbf{x}}\right\|_{2}}{\left\|{\mathbf{x}}-a\,{\mathbf{x}}^{*}+b\right\|_{2}},
\end{equation}
where a higher RSNR corresponds to a better reconstruction.

\subsection{Implementation Details}\label{sec: Training}
We use an Intel i7-7820X (3.60GHz) CPU and an NVIDIA Titan X (Pascal) GPU. Pytorch 1.0.0 on Python 3.6 is used to implement our generative model\footnote{We shall provide a link to the repository upon paper acceptance.}. The network is optimized until $n_\text{iter}=10,000$ with $B=1$ using Adam optimizer~\cite{kingma2014adam} of default setting and the learning rate of $10^{-3}$. 
\subsection{Architectures}\label{sec: Architectures}
The mapping network $g_\phi$ is two consecutive fully connected layers of 512 hidden dimension with ReLU in between. It outputs $L=64$-dimensional latent vector, which is reshaped to $\left(8\times8\right)$ for the following generative network $h_\psi$ (Table~\ref{tab:arch}). 
The generative network consists of convolutional layers, batch normalization layers, ReLU, and nearest-neighbor interpolations. We apply zero-padding before convolution to let the size of the output mirror that of the input. At the last layer, ReLU is not used. 
The output has two channels because MRI images take complex values.

\begin{table}[!t]
\vspace{-2mm}
\centering
\begin{tabular}{l@{\hspace{1.5mm}}l|c}
\hline\hline
         & \textbf{Method}        & \textbf{RSNR (dB)}    \\\hline
\arch{}  & Back Projection (BP; zero-filled DFT)                      & 8.39           \\
\arch{}  & GRASP \cite{feng2014golden}                              & 24.08          \\ 
\arch{} & Straight line ($L=64$)                     &\textbf{ 26.51 }         \\\hline\hline 
\end{tabular}
\vspace{1mm}
\caption{Performance on the retrospective dataset for a single heart cycle.} 
\label{tab:singlecycle}
\vspace{-5mm}
\end{table}
\section{Design of the latent space}\label{sec: Analysis}
In this section, we analyze the individual components of our model and compare the performance with baselines. We first demonstrate the simplest setup that reconstructs a single heart cycle. We then move on to a more complicated dataset that has multiple heart cycles.

\subsection{Straight-Line Manifold for a Single Heart Cycle}
A straight-line manifold can help the network to encode the temporal variations of images. To implement it, we first sample $\M z_0,\, \M z_{K-1}\sim \Spc U(\mathbb{R}^{L})$. Then, the intermediate $\M z_k$ are obtained by linear interpolation. 
This yields a straight-line manifold that simply joins the end points as
\begin{align}
    \M z_k=(1-\alpha_k)\, \M z_0 +\alpha_k\,\M z_{K-1},
\end{align}
where $\alpha_k=k/{(K-1)}$. 

Although simple, this configuration already outperforms the other baseline methods and successfully reconstructs the dynamics for a single cycle dataset (\Tref{tab:singlecycle}).  

\subsection{Manifolds for Multiple Heart Cycles}
In practice, the measurements generally span several heart cycles. To better exploit the fact that the cardiac movement has a quasi-periodic behavior, it is of interest to explore more sophisticated manifolds. 
\begin{itemize}
\item\textbf{Segmented Line}. We first sample $p+1=14$ landmarks $\{{\mathbf{z}}^{(\tau)}\}_{\tau\in[0\ldots13]}\sim \Spc U(\mathbb{R}^{L})$, where $p$ is the number of cardiac periods. We generate a set of equispaced intermediate ${\mathbf{z}}_{k}$ of each segment by a linear combination of $\mathbf{z}^{\tau}$ and ${\mathbf{z}}^{(\tau+1)},~\forall \tau\in [0\ldots12]$.

\item \textbf{Circles}. Let $\M z_k=[z_1^{(k)}, z_2^{(k)}, \M z_{\rm slack}]\in\mathbb{R}^L$ and $\M z_{\rm slack} \sim \Spc U(\mathbb{R}^{L-2})$. The first two coordinates $(z_1^{(k)},z_2^{(k)})$ are points from a unit circle with $p$ cycles. The slack coordinates do not depend on $k$. Thus, we have that 
\begin{align}
    \M z_k= \left[\cos(\frac{2\pi\, p\,k}{(K-1)}),\sin(\frac{2\pi\, p\,k}{(K-1)} ), \M z_{\rm slack} \right].
\end{align}
\item \textbf{Helix.} Similar to ``Circles", the first two coordinates of $\M z_k$ are points from a unit circle with $p$ cycles. The slack coordinates $\M z_{\rm slack} \sim \Spc U(\mathbb{R}^{L-2})$ are now scaled by $\frac{k}{(K-1)}$, learning to
\begin{align}\label{eqn:helix}
    \M z_k= \left[\cos(\frac{2\pi\, p\,k}{(K-1)}),\sin(\frac{2\pi\, p\,k}{(K-1)} ),  \frac{k\,\M z_{\rm slack}}{(K-1)} \right].
\end{align}
\end{itemize}


\begin{figure*}[!ht]
\vspace{-2mm}
\centering
\includegraphics[width=\linewidth]{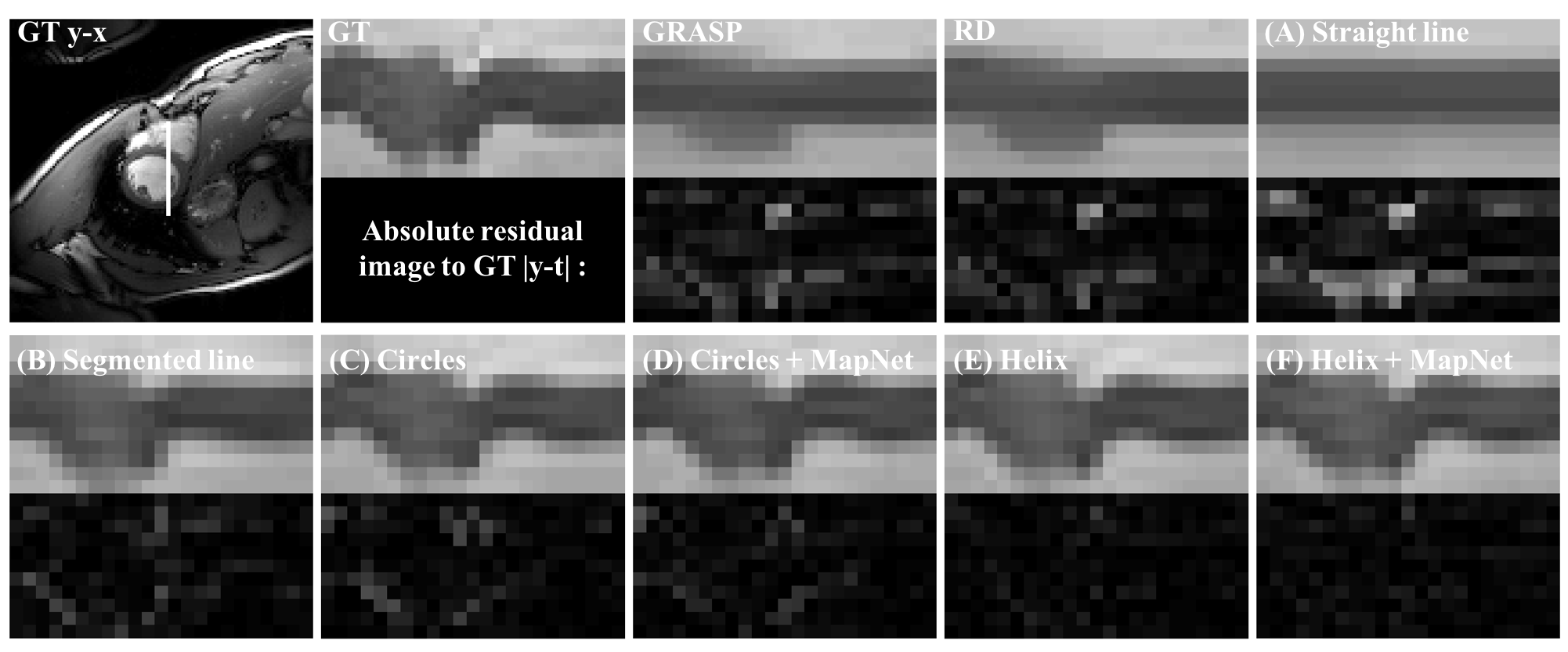}
\caption{\label{fig:retro} Visual comparison of reconstructed (y-t) images using the baseline methods and ours with each configuration in \Tref{tab:ablation}. The reconstructed images from fully sampled Cartesian trajectories are used as a ground-truth. 
A white line at the heart region indicates the cross section that is visualized. 
Here, for simulating RD~\cite{yerly2016coronary,chaptinel2017fetal}, we reorder the spokes of each frame from 13 periods resulting in 169 spokes per frame. 
($13\text{ periods}\times13\text{ spokes/frame}=169\text{ spokes/frame}$).
For better comparison, the residual images to the ground-truth are provided in the lower panels. }
\end{figure*}

\begin{table}[!t]
\vspace{1mm}
\centering
\begin{tabular}{l@{\hspace{1.5mm}}l|c}
\hline\hline
         & \textbf{Method}        & \textbf{RSNR (dB)}    \\\hline
\arch{}  & Back Projection (BP)                      & 8.4598           \\
\arch{}  & GRASP \cite{feng2014golden}                              & 24.2123          \\ 
\arch{}  & Reordered method (RD)~\cite{yerly2016coronary,chaptinel2017fetal}                               & 25.0364          \\
\arch{} & Straight line ($L=64$)                   & 20.55 $\pm$ 0.09          \\  
\arch{} & Segmented line ($L=64$)                  & 25.94 $\pm$ 0.16          \\
\arch{} & Circles ($L=64$)                         & 27.13 $\pm$ 0.15          \\
\arch{} & Circles ($L=2$) + MapNet ($L=64$)            & 27.52 $\pm$ 0.11          \\
\arch{} & Helix ($L=64$)                           & 27.78 $\pm$ 0.07          \\
\arch{} & \textbf{Helix ($L=3$) + MapNet ($L=64$)}     & \textbf{28.05 $\pm$ 0.04}         \\\hline\hline        
\end{tabular}
\vspace{1mm}
\caption{Performance on the Retrospective dataset for multiple heart cycles. Averaged RSNR over three runs and their standard deviations for several CNN latent space designs.} 
\label{tab:ablation}
\end{table}

\medskip
\noindent\textbf{Effect of the Manifolds.}
In \Fref{fig:retro}, we show the reconstructed (y-t) images of the cross section that is denoted by a white line in GT (y-x) image\footnote{For display purposes, we show only one cycle of our cross section.}. 
When we use a straight-line manifold, the network fails to capture the heart movement and outputs the same static image over all frames. This is natural since most of the pixels are static and the dynamic parts are localized in a small area. Thus, the network easily finds a local minimum that corresponds to an image that remains constant over all frames. However, as soon as we switch to ``periodic-like" manifold designs, the network starts to reconstruct the movement (\Tref{tab:ablation}). For example, when we use a line with 13 segments as an input, the performance is better than the RD that uses the same information. Using circles with 13 repetitions as input manifold, we improve even further. However, the helix input manifold gives the best performance among the others without MapNet because the heartbeat is a quasi-periodic signal. 

\medskip
\noindent\textbf{Effect of the Mapping Network.}
In addition to the choice of its manifold, our method has another design choice: its mapping network. 
By introducing MapNet, the network can adapt its input manifold to a given dataset, which allows us to further improve the reconstruction (\Tref{tab:ablation}). 
This can be clearly seen in the t-SNE visualization of the mapped latent space (\Fref{fig:tsne}), which we discuss in \S\ref{sec: Discussion}. 

In summary, 
our analysis shows that a careful design of the manifold and the use of a mapping network are both necessary to achieve the best performance. 
Based on these, from now on, we use `Helix+MapNet' as our default setup.

\section{Results}\label{sec: Results}

We first show results on the retrospective dataset, where the desired behaviors of the reconstruction methods are well-defined. 
We then illustrate on the fetal cardiac dataset that the observations extend well to a real scenario. 


\begin{figure*}[!t]
\vspace{-2mm}
\centering
\includegraphics[width=\linewidth]{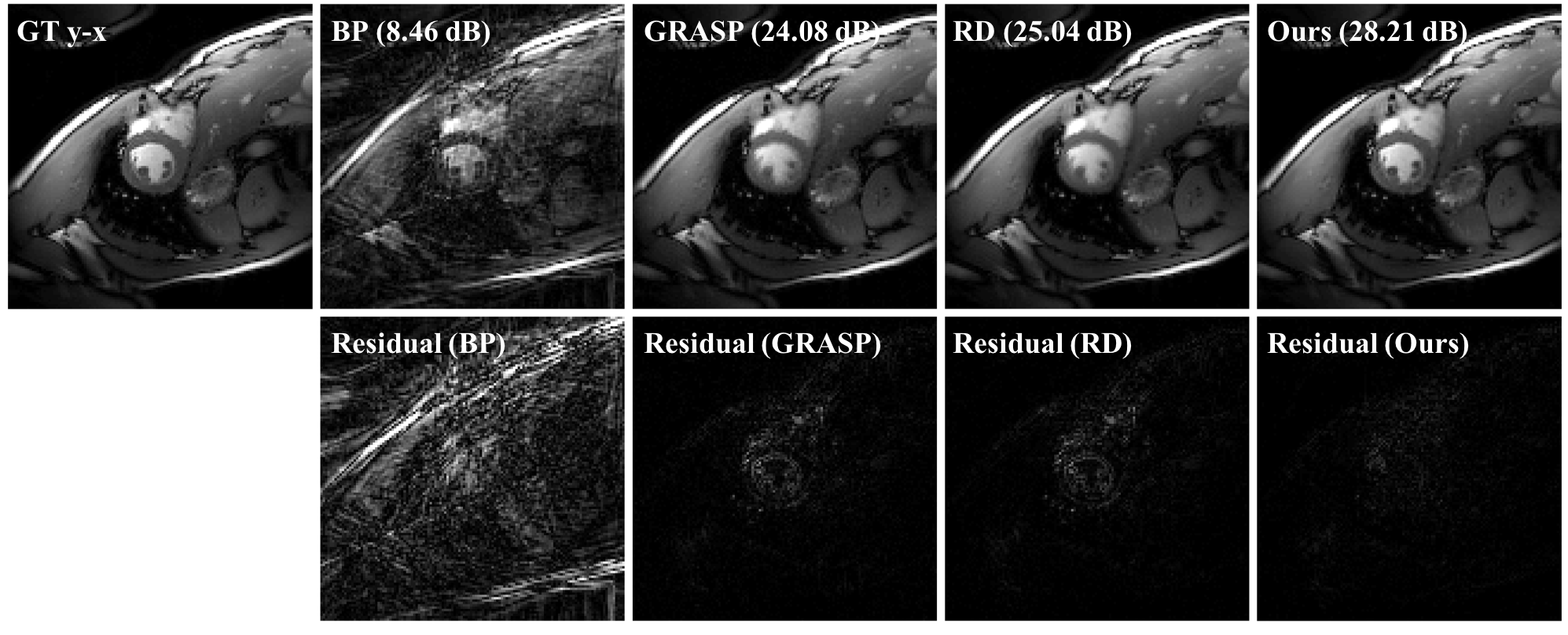}
\caption{\label{fig:retro2}
Visual comparison of reconstructed (y-x) images using BP, GRASP \cite{feng2014golden}, RD~\cite{yerly2016coronary,chaptinel2017fetal}, and our method (Helix + MapNet). The reconstructed images from fully sampled Cartesian trajectories are used as a ground-truth. Here, the RSNR value is for the single frame that is visualized. 
To simulate RD, we reorder the spokes of each frame from 13 periods resulting in 169 spokes per frame 
($13\text{ periods}\times13\text{ spokes/frame}=169\text{ spokes/frame}$). For better comparison, the residual images to the ground-truth are given together.}
\end{figure*}
\begin{figure*}[!t]
\vspace{-2mm}
\centering
\includegraphics[width=\linewidth]{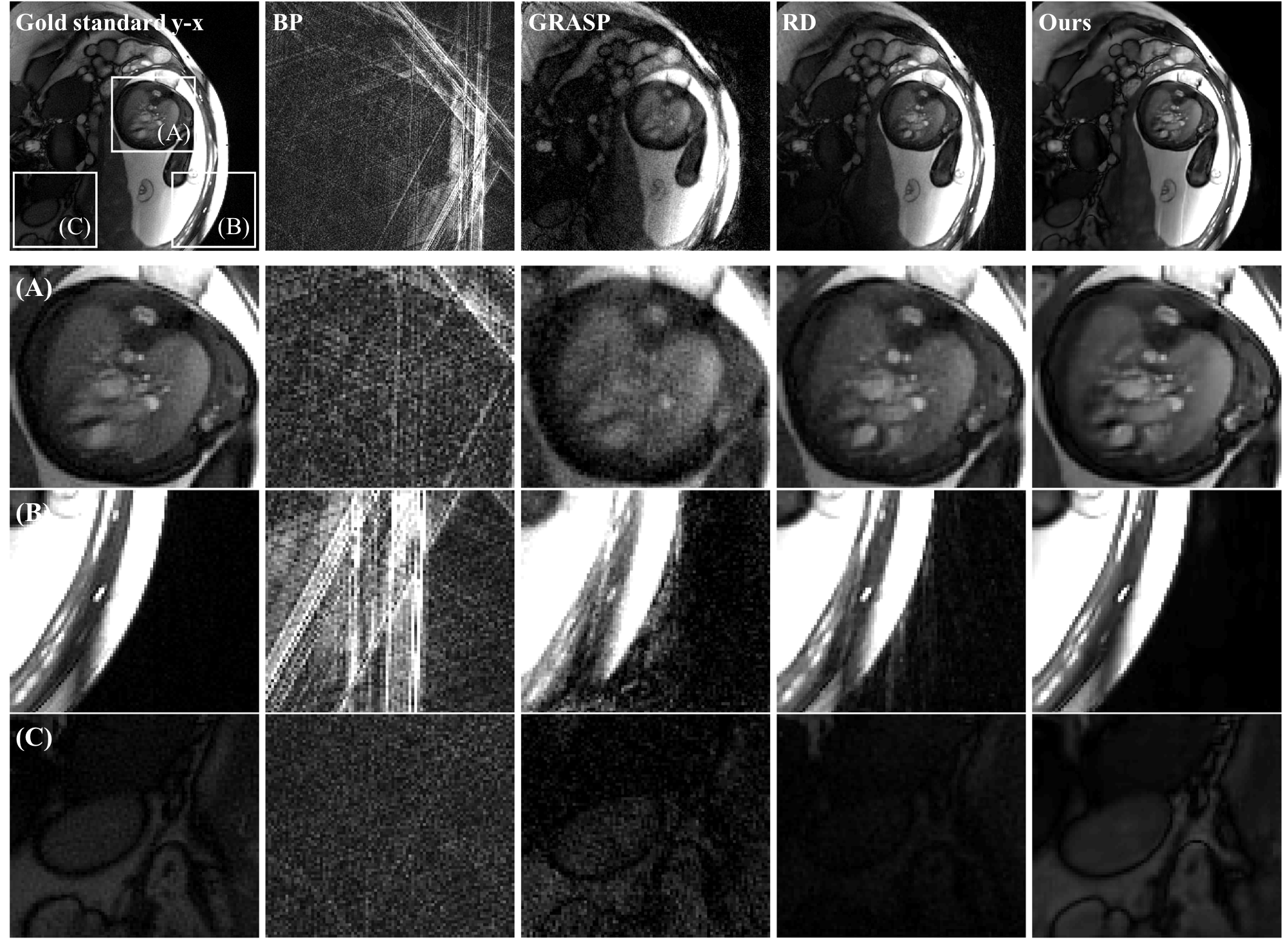}
\caption{\label{fig:chuv1}
Visual comparison of reconstructed fetal hearts; (y-x) images. The gold standard is reconstructed from the simultaneous use of all time frames (first column). Top row is the results of BP, GRASP \cite{feng2014golden}, RD~\cite{yerly2016coronary,chaptinel2017fetal}, and our method (Helix + MapNet). Bottom three rows (\arch{a}-\arch{c}) show magnified view.
}
\end{figure*}

\begin{figure*}[!t]
\vspace{-2mm}
\centering
\includegraphics[width=\linewidth]{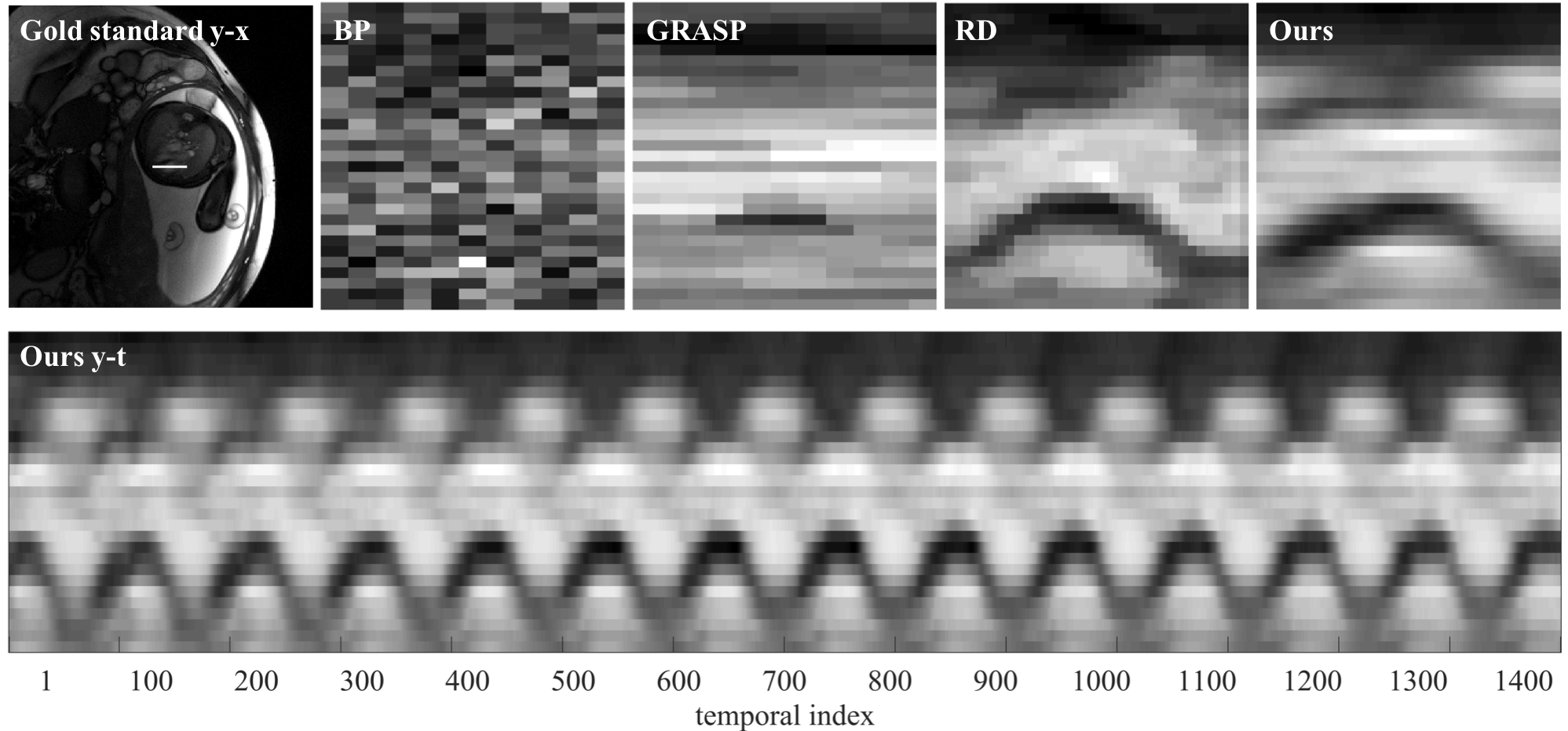}
\caption{\label{fig:chuv1_yt}
Visual comparison of reconstructed fetal hearts; (y-t) images. A white line in the gold standard indicates the cross section that is visualized. Top row is the results of BP, GRASP \cite{feng2014golden}, RD~\cite{yerly2016coronary,chaptinel2017fetal}, and our method (Helix + MapNet). Bottom row is the series of (y-t) cross sections of our reconstruction.}
\end{figure*}

\subsection{Retrospective Dataset: Multiple Heart Cycles}\label{sec:retro}

The benefits of our method are evident in both the (y-t) view (\Fref{fig:retro}) and (y-x) view (\Fref{fig:retro2}) of each frame. 
In \Fref{fig:retro}, both GRASP and RD reconstruct the movement of the heart. RD shows better performance than GRASP, which was expected because it takes advantage of the period information that is estimated while reordering the frames. However, as can be seen in the residuals, GRASP and RD show significant errors in the reconstruction of the dynamics. 
In the (y-x) view of \Fref{fig:retro2}, GRASP leads to blurring artifacts, while the residual image reveals errors around the wall of the heart in both GRASP and RD reconstructions. By contrast, our method gives better results with fewer artifacts. 
\subsection{Fetal Cardiac Dataset}\label{sec:fetal_heart}
Having demonstrated the superior behavior of our method on the retrospective dataset, we now assess our model on real data. In the absence of ground-truth, we shall take the static image that is generated from all spokes as pseudo-gold standard---note that  it is of high quality only in the regions that are not moving. 

Like in the retrospective experiments, 
both GRASP and RD are able to reconstruct multiple cardiac phases. RD gives better reconstructions, especially in the dynamic region (\Fref{fig:chuv1} (A)). However, RD shows a spurious artifact at the edge area (\Fref{fig:chuv1} (B)) and fails to find the detailed structures of the static background (\Fref{fig:chuv1} (C)). By contrast, our method produces better-resolved features in both dynamic and static areas (particularly for the hyperintense dot-like structures in \Fref{fig:chuv1} (A)), while it does not suffer from artifacts at the edges and recovers the low-intensity background areas as well (\Fref{fig:chuv1} (C)). 



In Figure~\ref{fig:chuv1_yt}, it is apparent that BP completely fails in capturing the fetal cardiac beats. The GRASP reconstruction is less noisy but still far from satisfactory. RD fares better; unfortunately, its reordering process can lead it to superpose in the same frame spokes that belong to different phases of the cardiac cycle. By contrast, our method reconstructs each frame with data from just a few neighboring spokes, thus avoiding the mingling of different cycles. The reconstructed systolic phase captures the true motion of the heart better. The cross section from our method is similar to that of RD but the motion is smoother in our case, which is the expected behavior of a beating heart.

We provide in Figure~\ref{fig:chuv1_yt} (Bottom row) our whole reconstructed sequence of cardiac cycles. 
The quasi-periodicity of the cardiac motion is clearly visible along the temporal axis, while motion variations can still be discerned from cycle to cycle. Note that this is a unique benefit of our method that the other algorithms cannot provide. 


\section{Discussion}\label{sec: Discussion}

\begin{figure*}[t!]
\vspace{-2mm}
\centering
\includegraphics[width=\linewidth]{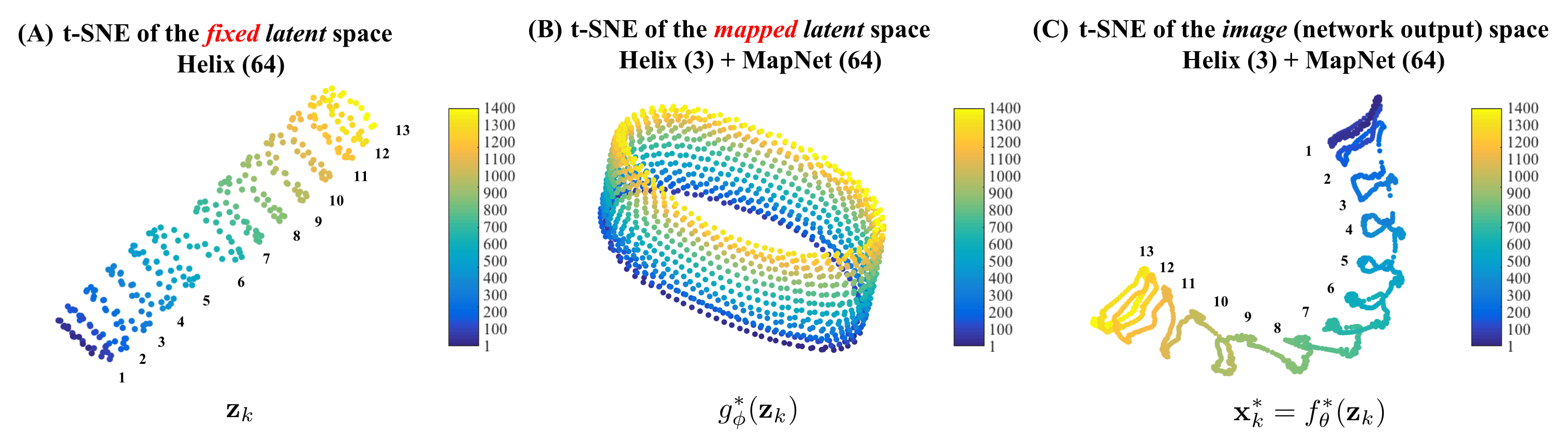}
\caption{\label{fig:tsne} Visualization of three-dimensional t-SNE embeddings of (A) the fixed latent space of a helix in 64-dimension (`Helix'); (B) the mapped latent space by the mapping network in 64-dimension (`Helix + MapNet'); and (C) its corresponding image space in $(256\times 256)$ dimension. Here, the temporal index is color-coded ($1,400\text{ frames}$). There are approximately 13 cycles of heart motion, which are also clearly seen in the embedded helix of the reconstructed images. 
}
\end{figure*}

\subsection{t-SNE Visualization of the Latent and Image Spaces}\label{sec: t-sne}
To assess the extent of structural change as a function of time, we used t-stochastic neighborhood embeddings (t-SNE)~\cite{maaten2008visualizing} which capture the underlying manifold by projecting the high-dimensional entities onto a three-dimensional space. 
In \Fref{fig:tsne} (A), we show the t-SNE result of the original manifold when the variables are generated according to \eqref{eqn:helix} with $L=64$ and $p=13$. Unsurprisingly, this recovers a helix with 13 cycles. 
In \Fref{fig:tsne} (B), we show the embedding of the 64-dimensional mapped variables $\{g_{\phi^*}(\M z_k)\}_{k=0}^{K-1}$, where the $\M z_k$ are generated according to \eqref{eqn:helix} with $L=3$ and $p=13$. Again, we recover a helical geometry with 13 periods, although the height of the helix is now shortened---the first and the last cycles become closer. This shows that MapNet successfully recovers some similarity between the different cycles which, in turn, translates into better reconstructions. It warps the given manifold in adaptive fashion, while retaining the prior information that we inject via the manifold geometry. 
As shown in \S\ref{sec: Analysis}, this design (Helix+MapNet) outperforms the `Helix' with $L=64$ and $p=13$ which is fed directly to the vanilla CNN. 
In \Fref{fig:tsne} (C), we display the projected manifold of the $(256\times 256)$-dimensional reconstructed images. 
It shows a helical structure with 13 local folds, each of which corresponds to a single cycle of the cardiac motion. 
This also shows that the quasi-periodic characteristic of the data is well represented by the network. 
\subsection{Benefits of Our Approach}\label{Contributions}

\noindent\textbf{Continuous Dynamic Reconstruction.} 
One major benefit of our approach is that it lets us reconstruct temporally continuous dynamic images. We showed that the network $f_{\boldsymbol\theta^*}$ successfully captures the underlying nonlinear dynamics of the image manifold, and the input variable $\M z_k$ lets us reconstruct the image at the corresponding time stamp (\Fref{fig:tsne}).  Because our method represents images as a learned parametric function $f_{\boldsymbol\theta^*}$, we can recover nontrivial intra-frame images by navigating between two consecutive input variables, which would not be possible with other standard interpolation methods such as temporal bilinear interpolation. 


\medskip

\noindent\textbf{Memory Savings.}\label{sec: Memory Savings}
In the methods based on compressed sensing (CS), the gradient updates of the iterative optimization process necessitate memory that is large enough to hold the target reconstruction volume. For example, the reconstruction of $5,\!000$ frames with spatial size $\left(256\times256\right)$ would need one to handle data of size $\left(256\times256\times5,000\right)$, which demands for over a gigabyte of memory.
Our approach, by contrast, requires much less memory. It optimizes the neural network using batches, which requires the simultaneous handling of only those frames that correspond to the batch size. In short, the fact that our proposed approach handles few 2D images whereas CS handles a 2D+t extended sequence leads to substantial savings, particularly for golden-angle dynamic MRI with many frames. In our approach, we only store a 2D generative model; for example, its memory demands for the spatial size $\left(256\times256\right)$ are about half-a-dozen megabytes. This cost is negligible compared to that of the CS approach.

\medskip
\noindent\textbf{Efficient Reconstruction.}
Our model visits each frame about seven times during training (10,000 iterations / 1400 frames $\approx 7$ ), while GRASP sees all frames during the entire iterations (24 outer iterations). 
Regarding the execution time, the major bottleneck of our method is the slow forward model. It depends on the NuFFT package which, in its current implementation, does not benefit from a GPU and is a major cause for slowdown. Indeed, NuFFT takes 47 \% of the entire running time of our algorithm per each iteration; the average processing time for 100 repetitions is 6.55 s for back and forth NuFFTs, and 3.08 s for the remaining parts. With a more efficient implementation, our algorithm could be substantially accelerated.

Because our model is fully automated, it leads to a simpler optimization task with fewer hyperparameters than the conventional methods. For instance, k-t SENSE requires three interdependent hyperparameters whose optimal values are found only after some substantial grid-search effort, while the two hyperparameters of our approach are easier to interpret since they trivially consist of just an initial learning rate, along with a number of iterations. 

\section{Conclusion}
In this paper, we proposed an unsupervised deep-learning-based algorithm for dynamic MRI reconstruction that provides high spatial resolution with access to the sub-frame---or even continuous---temporal control of dynamic images. By designing a one-dimensional manifold, combined with the mapping network, our generative network model fully exploits the representation power of the network as well as its structural priors. Our study showed that the proposed method successfully reconstructs dynamic MRI in an end-to-end manner and outperforms the state-of-the-art CS approaches by 3.8 dB. To the best of our knowledge, this is the first unsupervised-learning approach in accelerated dynamic MRI. 


\section*{Acknowledgements}
The authors thank Prof.\ Jong Chul Ye at KAIST for providing the bSSFP  cardiac MRI k-space dataset (retrospective dataset).




\bibliographystyle{IEEEtran}
\bibliography{sample}
\end{document}